\documentclass[aps,prd,10pt,notitlepage, nofootinbib,
superscriptaddress,showkeys,showpacs]{revtex4-1}
\usepackage{graphicx}

\usepackage{amstext,amsmath,amssymb,amsfonts,bbm}
\usepackage{epsfig}
\usepackage{hyperref}
\usepackage{amsthm}
\usepackage{color}

\usepackage{graphicx}
\usepackage{latexsym}
\usepackage{cancel}

\usepackage{mathtools}

\newcommand{\bea}{\begin{eqnarray}}	
\newcommand{\eea}{\end{eqnarray}}
\newcommand{\be}{\begin{equation}}	
\newcommand{\ee}{\end{equation}}
\newcommand{\beq}{\begin{equation}}	
\newcommand{\eeq}{\end{equation}}

\newcommand{\C}{{\mathbb C}}
\newcommand{\Z}{ {\mathbb Z} } 
 
\newcommand{\N}{ {\mathbb N} } 
\newcommand{\cG}{\mathcal{G}}
 
 \newcommand{\cV}{\mathcal{V}}
 
 \newcommand{\cL}{\mathcal{L}}

\newcommand{\cF}{ { \mathcal{ F } } }

\newcommand{\Sym}{ {\rm Sym} }

\newcommand{\cexG}{\mathcal G_{\rm{color}}}

\newcommand{\bG}{{\partial\mathcal G}}
\newcommand{\bJ}{ J_{\partial} }
\newcommand{\tJ}{{\widetilde{J}}}

\newcommand{\R}{\mathbb{R}} 
\newcommand{\Tr}{{\rm Tr}}

\newcommand{\rmd}{ {\rm d} }

\newcommand{\bdel}{{\boldsymbol{\delta}}}
\newcommand{\bmu}{\boldsymbol{\mu}}

\newtheorem{theorem}{Theorem}
\newtheorem{proposition}{Proposition}

\newcommand{\inter}{{\rm int\,}}
\newcommand{\ext}{{\rm ext\,}}

\begin{document}

\title{ A power counting theorem for a
 $p^{2a}\phi^4$ tensorial group field theory}

\author{Joseph Ben Geloun}\email{jbengeloun@aei.mpg.de}

\affiliation{Max Planck Institute for Gravitational Physics, Albert Einstein Institute\\
 Am M\"uhlenberg 1, 14476, Potsdam, Germany }

\affiliation{International Chair in Mathematical Physics and Applications, 
ICMPA-UNESCO Chair, 072BP50, Cotonou, Rep. of Benin }

\date{\small\today}

\

\

\begin{abstract}

We introduce a tensorial group field theory endowed with 
 weighted interaction terms of the form $p^{2a} \phi^4$.
The model can be seen as a field theory over $d=3,4$ copies
of $U(1)$ where formal powers of Laplacian operators, namely $\Delta^{a}$, $a>0$, act on tensorial $\phi^4$-interactions producing, after Fourier transform,
 $p^{2a}\phi^4$ interactions. Using multi-scale analysis, 
we provide a power counting theorem for this type of models. 
A new quantity depending on the incidence matrix between 
vertices and faces of Feynman graphs is invoked in the degree of 
divergence of amplitudes. As a result,  generally, 
the divergence degree is enhanced compared to the divergence
degree of models without weighted vertices. The subleading
terms in the partition function of the $\phi^4$ tensorial models  become, in some cases, the dominant ones in the $p^{2a}\phi^4$ models. Finally, we explore 
sufficient conditions on the parameter $a$ yielding a list of potentially super-renormalizable $p^{2a}\phi^4$ models.

\medskip
\noindent Pacs numbers: 11.10.Gh, 04.60.-m, 02.10.Ox
 ICMPA/MPA/2015/04

 \bigskip 
\begin{center}

{\it For Vincent Rivasseau, in this year of his 60th birthday}

\medskip

{\it ``Je n'enseigne pas, je raconte.'' de Montaigne.}

\medskip

{\it ``Behandle die Menschen so, 
als w\"aren sie, was sie sein sollten,\\
und du hilfst ihnen zu werden,
was sie sein k\"onnen.'' Goethe.}
\end{center}

\end{abstract}

\maketitle

\section{Introduction} 

Tensorial Group Field Theory (TGFT) \cite{Rivasseau:2011hm} is a field theoretical formulation of tensor models \cite{tensor} which has also strong ties with Group Field Theory \cite{GFTreviews}, placing at its heart the combinatorial duality between Feynman graphs and simplicial manifolds. 
It is also fair to mention that they all emanate from the study of matrix models \cite{Di Francesco:1993nw}, a well-known success addressing gravity in 2D, 
and all pertain to dedicated efforts for defining a discrete-to-continuum scenario for gravity. 

Recently, the study of these tensor models has acknowledged a strong
revival because of the discovery of a well behaved sub-class of these
models, the colored ones \cite{coloured}. Colored tensor models
have been intensively studied  because they possess an interesting
1/N expansion \`a la 't Hooft \cite{largeN}. Such a 1/N expansion, in the 2D case, was a key aspect leading to the understanding of phase transitions in matrix models  \cite{Di Francesco:1993nw}.   
Concerning colored tensor models, the 1/N expansion also allows
to analytically prove that there exists indeed phase transition in such models
\cite{criticalTensor}. Nevertheless, one shows that the new phase  
does not describe geometries of the expected type but a 
singular branched polymeric one \cite{Gurau:2013cbh}. Because tensor models
have a lot more structure, it has been suggested the existence of 
multiple-scaling limits to examinate other regimes of parameters
which allow to incorporate a wider class of graphs (including subleading) such that
the critical behavior of tensor models could be improved \cite{Dartois:2013sra,Bonzom:2014oua}. This is clearly a greater
challenge which is still under investigation. 
 It might therefore be appropriate to ask ``Is the branched polymer phase not  simply the fate of tensor models?''.\footnote{This question was raised by Thibault Damour, during the conference ``Quantum Gravity in Paris,'' IHES \& LPT Orsay, March, 2014.} 
A question of this kind cannot find a short and authoritative
answer, at least, within a short period of time. What is however certain is that, as a  physically motivated mathematical framework, tensor models offer enough freedom to be enriched, hopefully towards reaching 
their initial goals. In the present work, going in that direction, we will discuss a class
of tensor models which are susceptible to gain ground upon the above undesirable geometrical limit.

Supplementing the statistical point of view, colored tensor models find a field theory formulation that one calls TGFT.\footnote{We must stress that,
in this work, we will interchangeably use TGFT and tensorial models.} The idea here is to understand in a field theory language the Renormalization Group analysis of tensor models and their ensuing flow of coupling constants. Several renormalizable models have been analyzed perturbatively 
and proved renormalizable and their Renormalization Group flow has been studied \cite{bgr}-\cite{Benedetti:2014qsa}.
More recently, as a prominent way for addressing the issue of critical 
phenomena, the Functional Renormalization Group approach 
has been investigated for tensorial models  \cite{Benedetti:2014qsa}. 
The RG flow of the simplest rank 3 TGFT over the torus $U(1)^3$ is determined as
a non-autonomous system of $\beta$-function. The explicit appearance of the IR cut-off in the RG equations results from the existence of a hidden scale which is the radius of the manifold and the nonlocal 
feature of the interactions. At large or small radius limit, one is able to set up a proper notion
of dimensionless couplings and then infers the existence of fixed
points in the flow. The presence of a  non-Gaussian fixed point supports
the existence of a phase transition again. 
As in the usual  scalar field theory, one suggests that these phases must be related 
to a symmetric phase and broken (or condensate) phase. Such a program 
must be pursued to clarify this issue.

In order to make clear the objectives of this work, let us detail
some aspects of these TGFTs. 
The expansion of the partition function of the colored tensor 
models admits a 1/N expansion in a new parameter 
called the degree of a colored tensor graph \cite{largeN}. 
The degree is a combinatorial quantity replacing the
genus, from 2D to arbitrary $d$ dimensional colored simplicial 
complex.  
In that expansion, the leading terms are special 
class of graphs called ``melons'' \cite{criticalTensor}, or ``melonic graphs,''
which are dual to peculiar sphere triangulations in any dimension $d$. 
These graphs satisfy a kind of planarity condition 
(similar to planarity for ribbon graphs of matrix models)
and are spanned by a recursive rule 
by inserting two-point graphs
onto lines of two-point graphs.
Among the ideas to escape from the branch 
polymer phase in tensor models, one proposal is to enhance the contribution of graphs 
of the non-melonic type. Such a program has been recently investigated
in the statistical framework \cite{Bonzom:2012wa,Bonzom:2015axa}. Our following
scheme inspires from these but, it is formulated in a radically different approach which is the field theory one. 

Given a rank $d$ complex tensor $\phi_{\bf p}$, where ${\bf p}=(p_i)$ is a multi-index. 
The known renormalizable TGFT actions are defined using  \cite{rankd}:

- a kinetic term of the form $\sum_{\bf p}\bar\phi_{\bf p}(|{\bf p}|^{2b}+\mu)\phi_{\bf p}$, where $b$ is a positive parameter and $\mu$ a mass coupling, $|{\bf p}|^{2b}=\sum_i |p_i|^{2b}$; this choice is reminiscent of a sum of powers of 
eigenvalues of a Laplacian, if one looks at $\phi_{\bf p}$ as the Fourier components
of a field;  

- nonlocal interactions of the melonic type obtained by convoluting an even 
number of tensors that we write $\Tr_{2n}( \phi ^{2n})$. 

In this work, we restrict our attention to the ranks $d=3,4$, and
in addition to the above parts, we introduce another vertex operator with a momentum weight obtained by convoluting $2n$ tensors, using 
the same pattern of contraction of melonic vertices but modified 
with an index dependent kernel. We write the new vertex in the suggestive form $\Tr_{4}(p^{2a}\phi^4)$. In direct space, for the value $a=1$, such terms have clear-cut meaning: these are obtained by letting act Laplacians on $2n$ fields convoluted. For arbitrary $a$, these vertices can be certainly written in the momentum space.  In a pictorial way, we propose to put a weight, $p^{2a}$,  to the melonic vertex (attached
to some strands of the vertex) in a way to generate, at the quantum and perturbative level, non-melonic graphs with enhanced power counting. 

By multi-scale analysis \cite{Rivasseau:1991ub}, we identify a power counting theorem for the $p^{2a}\phi^4$-model. As expected, non-melonic contributions are enhanced and these can be 
even more divergent than melonic ones. To the degree of divergence of any 
graph obtained in previous analysis \cite{rankd},  we must now add a new quantity. This combinatorial object is obtained from the optimization of the integrations
of internal momenta (equivalent to a momentum routine in ordinary QFT) 
which involves  a new incidence matrix between vertices and faces or strands of the graph. We recall that a power counting theorem 
is essential in understanding a renormalization analysis. We do not perform this  procedure here but simply undertake the first steps. Scrutinizing the degree of divergence of any graph in 
ranks $d=3$ and 4, and at maximal valence of the vertex $\phi^4$, a list of sufficient conditions on the parameter $a$ strongly suggests the existence of super-renormalizable  $p^{2a}\phi^4$ models. 
The formalism  easily extends in any rank $d$ with not much 
effort.

The paper is organized in the following way: the next section 
reviews the construction of tensorial models and presents
the new vertices. Devoted to 
the perturbative analysis at all orders, Section \ref{sect:perturb} starts
by the quantum model and its amplitudes, sets up the multi-scale
analysis, leads to our main result, namely the
power counting Theorem \ref{theo:pc} and finally discusses
potentially renormalizable models. We give a  conclusion 
of this work in Section \ref{ccl} and the paper closes with an appendix 
providing a worked out example of the optimization procedure
on which relies the multi-scale analysis.

\section{Models}
\label{sect:actio}

Consider a rank $d$ complex  tensor $\phi_{\bf P}$, with ${\bf P}=(p_1,p_2,\dots,p_d)$ a multi-index, and denote $\bar\phi_{\bf P}$ its complex conjugate. The indices $p_k$ can be
chosen of several types. For simplicity, 
in this work, we consider that these are integers: $p_k \in \Z$. This choice can be certainly motivated from a field theory point of view: introducing a complex function $\phi: U(1)^{d} \to \C$,
$\phi_{\bf P}$ define nothing but the Fourier components of such a field. 
Hence, the development as found hereafter could be translated in a field
theory language on a compact space like the $d$-torus (up to subtleties that we will give precisions on) and this is also the reason why, in most of our study, we regard 
 ${\bf P}$ as a momentum index.

An action $S$ of a tensorial model is built by convoluting several copies
of $\phi_{\bf P}$ and $\bar\phi_{\bf P}$ using kernels. $S$ is of the
general form 
\bea\label{eq:actiond}
&&
S[\bar\phi,\phi]=\Tr_2 (\bar\phi \cdot K \cdot \phi) 
+ \mu\, \Tr_2 (\phi^2) + S^{\inter}[\bar\phi,\phi]\,, 
\cr\cr
&&
\Tr_2 (\bar\phi \cdot K \cdot \phi) =
\sum_{{\bf P}, \, {\bf P}'} \bar\phi_{{\bf P}} \, K({\bf P};{\bf P}')\, \phi_{{\bf P}' } \,, 
\qquad 
\Tr_{2}(\phi^2) = \sum_{{\bf P}} \bar\phi_{{\bf P}}\phi_{{\bf P}}\,, 
 \cr\cr
&& S^{\inter}[\bar\phi,\phi]=  \sum_{n_b} 
\lambda_{n_b}  \Tr_{n_b}(\bar\phi^{n_b}\cdot \cV_{n_b}\cdot \phi^{n_b})\,,
\eea
where $\Tr_{n_b}$ are ``generalized traces'' over tensors,
$K$ and $\cV_{n_b}$ kernels to be specified, $\mu$ (mass) and $\lambda_{n_b}$
are coupling constants. Putting $\cV_{n_b}$ to an identity kernel, 
it must be pointed out that $\Tr_{n_b}$ are convolutions of tensors which 
generate unitary invariants \cite{Gurau:2011tj, Gurau:2012ix,Bonzom:2012hw}. 

Let us be more specific at this stage and characterize the kinetic term involving $K$ and the mass-like term $\Tr_{2}(\phi^2)$. We are interested in an action in rank $d$ with a kinetic term determined by 
\beq
\label{eq:3dkin}
K(\{p_i\};\{p_i'\}) = \bdel_{p_i,p_i'} ( \sum_{i=1}^d p^2_i)  \,,
\qquad 
\bdel_{p_i,p_i'} := \prod_{i=1}^d \delta_{p_i,p_i'} \,,
\qquad 
\Tr_{2}(\phi^2)= \sum_{p_i \in \Z} |\phi_{12\dots d} |^2 \,,
\eeq
where we use a compact notation $\phi_{12\dots d}:=\phi_{p_1,p_2,\dots, p_d}$.
The kernel $K$  is the sum of squared eigenvalues of $d$
Laplacian operators over the $d$ copies of $U(1)$. 
 Mostly, we will restrict our attention to the rank $d=3$ and $d=4$ cases. 

Focusing on the interaction part, given a parameter $a\in (0,\infty)$, the interaction terms are chosen such that 
\bea
\label{eq:3dinter}
&&
S^{\inter}[\bar\phi,\phi]=  \frac{\lambda}{2}\, \Tr_{4}(\phi^4)
+  \frac{\eta}{2}\, \Tr_{4}(p^{2a}\,\phi^4) \,, \cr\cr
&& 
 \Tr_{4}(\phi^4)
 :=  \Tr_{4;1} (\phi^4) + \Sym (1 \to 2 \to \dots \to d) 
\,, \cr\cr
&& 
 \Tr_{4}(p^{2a}\phi^4)
 :=  \Tr_{4;1} (p_1^{2a}\,\phi^4) + \Sym (1 \to 2 \to \dots \to d) \,, 
\eea
where $\lambda$ and $\eta$ are coupling constants and 
where the symbols $ \Tr_{4;1} (\phi^4)$, $\Tr_{4;1} (p_1^{2a}\,\phi^4)$ and
$\Sym$ must be now given a sense.  In rank $d=3$, the expression of the tensor traces
in $S^{\inter}$ have the explicit form
\bea
&& 
\Tr_{4;1}(\phi^4) = \sum_{p_i, p_i' \in \Z} 
\phi_{123} \,\bar\phi_{1'23} \,\phi_{1'2'3'} \,\bar\phi_{12'3'} \,, \cr\cr 
&&
\Tr_{4;1}(p_1^2\,\phi^4) = \sum_{p_i, p_i' \in \Z} 
\Big(p_1^{2a} + {p'}_1^{2a}\Big)\phi_{123} \,\bar\phi_{1'23} \,\phi_{1'2'3'} \,\bar\phi_{12'3'} \,, 
\label{firstchoice}
\eea
and, in rank $d=4$, these contractions easily generalize  using
the same cyclic pattern and by introducing
 another index (a graphical representation associated with these vertices will 
follow after introducing the quantum model and its Feynman rules, in the next section). The symbol $\Sym$ in \eqref{eq:3dinter} manifests the fact that we must add to the above terms colored symmetric ones. 
In the end, we shall write, in understandable and more compact notations,  
\beq
S^{\inter}[\bar\phi,\phi]=  \frac{1}{2}\,\Tr_{4}[( \lambda+ \eta p^{2a})
\,\phi^4]\,. 
\eeq
Note that the interaction kernels can be interpreted as well in the $U(1)$ formulation:
we add  formal operators $\Delta^{a}_s+\Delta^{a}_{s'}$ acting on the $\phi^4$ terms. The formal character of these operators is in the sense that, for arbitrary $a$, we regard these only through the momentum space. Hence, the coupling $\eta$ plays a role similar to a wave function renormalization
now associated with the interaction. This means that, when performing a renormalization procedure,  subleading contributions to the vertex
operators must be investigated. 
As one might realize from this point, the present theory space
becomes far richer than the usual unitary invariant potential ansatz 
where the vertices of the model do not have any 
momentum weight. 
A last thing must be noticed: in both class of models, with weighted vertices or not,  the interaction terms 
are nonlocal and the new combinatorics they generate provide them 
with a genuinely different renormalization analysis than ordinary quantum field theories.   

\section{Amplitudes and multiscale analysis}
\label{sect:perturb}

\begin{figure}[h]
 \centering
     \begin{minipage}{.7\textwidth}
\includegraphics[angle=0, width=11cm, height=1.8cm]{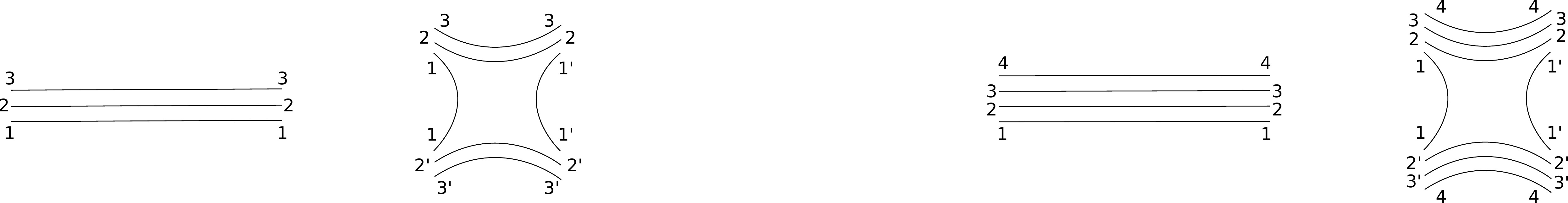} 
\vspace{1cm}
\caption{ {\small Rank $d=3$ and $4$ propagator (stranded) lines and vertices $\frac12\Tr_{4;1}[(\lambda + \eta p_1^{2a}) \phi^4]$.   }} 
\label{fig:4vertex}
\end{minipage}
\put(-305,-10){Rank $d=3$}
\put(-105,-10){Rank $d=4$}
\end{figure}

The quantum model associated with the above action 
\eqref{eq:actiond} with kinetic term \eqref{eq:3dkin} and interaction \eqref{eq:3dinter}
is determined by the partition function
\bea
Z  = \int d\nu_C(\bar\phi,\phi) \; e^{-S^{\inter}[\bar\phi,\phi]}\,, 
\eea
where $d\nu_C(\bar\phi,\phi)$ is a Gaussian field measure with covariance
given by the inverse of the kinetic term and this is
\bea
C(\{p_i\};\{p_i'\}) =\tilde{C}(\{p_i\})\, \bdel_{p_i,p_i'}\,,\qquad
\tilde{C}(\{p_s\})= \frac{1}{\sum_s p_s^2 + \mu}\,. 
\eea
At the graphical level the propagator is represented by a collection of $d$ segments
called strands (see Figure \ref{fig:4vertex}).  Dealing with the interaction, we have the following vertex kernel amplitude associated with \eqref{firstchoice}:
\beq
V_{4;s}(\{p_i\};\{p_i'\};\{p_i''\};\{p_i'''\}) = \frac12\Big(\lambda + \eta(p_s^{2a} + {p'}_s^{2a})\Big) \delta_{4;s}(\{p_i\};\{p_i'\};\{p_i''\};\{p_i'''\})\,,
\;\; s=1,2,\dots, d\,, 
\eeq
where the operator $\delta_{4;s}(-)$ is a product of Kronecker deltas identifying 
the different momenta according to the pattern given by 
the vertex $\Tr_{4;s}(\phi^4)$. We remark that $V_{4;s}$ 
has a color index. 
Graphically, the interaction  is represented
by stranded vertex (see $V_{4;s=1}$ or equivalently $\frac12\Tr_{4;1}[(\lambda + \eta p_1^{2a}) \phi^4]$ in rank $d=3$ and $4$, in Figure \ref{fig:4vertex}).  
Strictly speaking, one should
introduce two types of vertices, one for each coupling, $\lambda$ and
$\eta$. Such a requirement will be  mandatory when the renormalization
analysis will be carried out. Nevertheless, in the following, we are interested in a power counting theorem which can be achieved without any further distinction.  

As emphasized before, the vertex operator has now a weight. 
A way to circumvent this feature (we may note that this is not necessarily a negative one) is to bring back to the
propagator the total momentum dependence of the vertex.
This amounts, in our present case, to redefine a propagator kernel of the form
\beq
C'_{s}(\{p_i\};\{p_i'\}) = \frac{\sqrt{\frac{1}{2}(\lambda +  \eta(p_s^{2a} + {p'}_s^{2a}))}}{\sum_{s'} p_{s'}^2 + \mu}  \bdel_{p_i,p_i'}\,, \qquad 
s=1,2,\dots,d\,,
\label{wc}\eeq
which looks almost unusual and has now a manifest color index. By inspecting directly \eqref{wc},  we can infer  that, {\it a priori}, the large momenta analysis of the amplitudes will be governed by the presence of $\eta$-terms. 
We will give a precise statement about this in the following. 
This being mentioned, we will use the direct approach, i.e. considering a symmetric propagator and colored vertices which altogether make still a tractable model in the present situation.

\ 

\noindent{\bf Amplitudes.}
The graph amplitudes of the model have the following structure:
given a connected graph $\cG$ with set $\cV$ of vertices  (with $V=|\cV|$) and set $\cL$  of propagator
lines (with $L=|\cL|$), we write (in loose notations)
\beq\label{ampli}
A_{\cG} = \sum_{p_{v;s}} 
\prod_{l \in \cL } C_{l} (\{p_{v(l)}\}; \{p'_{v'(l)}\}) 
\prod_{v \in \cV} (-V_{4;v} (\{p_{v;s}\}))\,.
\eeq
This expression is similar to the ordinary field theory amplitudes 
where propagators $C_l$ have line indices $l$ and momentum arguments
$p_{v(l)}$  convoluted using vertex constraints $V_{4;v}$. 
The sum in \eqref{ampli} is performed over internal momenta $p_{v;s}$
appearing in the vertex operators $V_{4;v}$. 
Noting that the propagator and vertex operators are weighted discrete delta's, 
there is conservation of momenta along strands of the graph. 
In contrast with usual tensorial models where
the  amplitude factorizes along connected strands
called also faces of the graph \cite{bgr}, the amplitude \eqref{ampli}  here cannot 
be factorized but is a sum of strand-factorized terms. This new feature
is described in the next paragraph. 

\begin{figure}[h]
 \centering
     \begin{minipage}{.7\textwidth}
\includegraphics[angle=0, width=3.5cm, height=1.5cm]{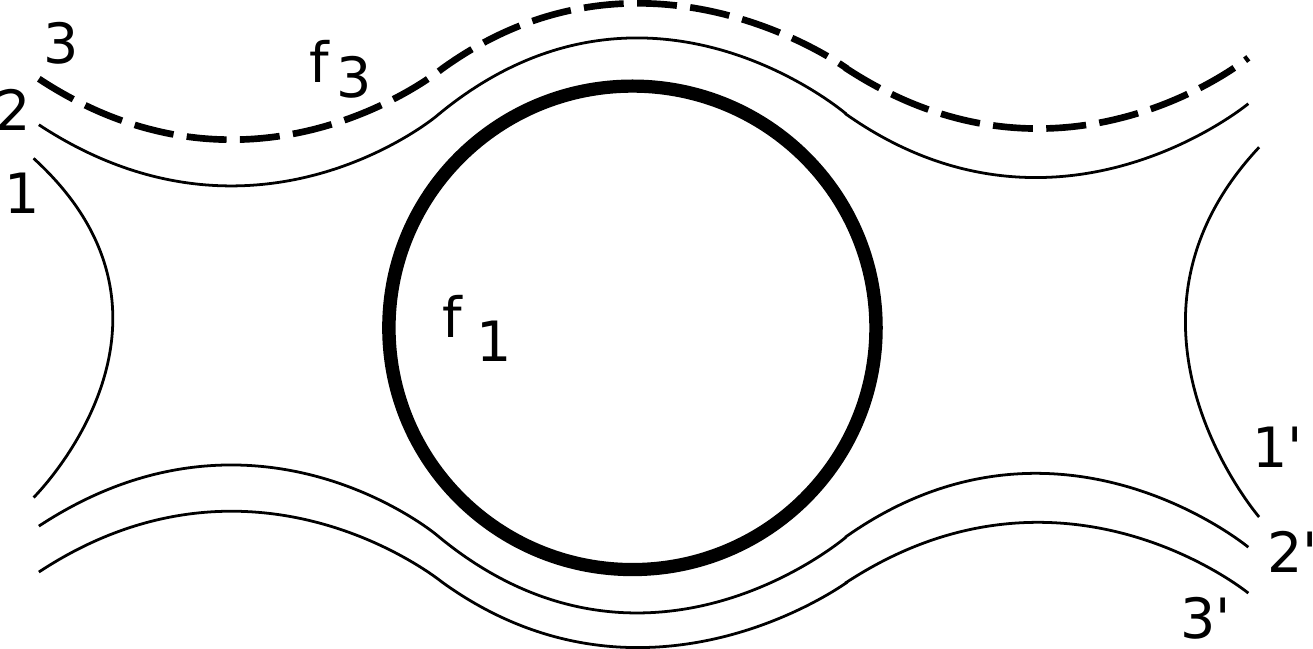} 
\caption{ {\small  A rank 3 4-point graph with an internal (closed) face ${\rm f}_1$ with color 1 (in bold). Any face different from $f_1$ is open or external, for instance, see the face ${\rm f}_3$ with color 3 (in dash). }} 
\label{fig:4ptfaces}
\end{minipage}
\end{figure}

Following strands in the tensor graphs, one defines one dimensional 
connected objects called ``faces'' (see Figure \ref{fig:4ptfaces}). 
One distinguishes two types of faces:
open ones, homeomorphic to lines, and closed or internal ones, homeomorphic to 
circles which are sometimes called loops.  The sets of external and internal faces are denoted by $\cF_{\ext}$ and $\cF_{\inter}$, respectively. A face in a graph has a colored index  $s=1,\dots, d,$ which  refers to a color index in the tensor. A face $f_s$ with color $s$ has a 
colored conserved momentum $p_{f_s}$ and passes through some vertices $v_{s}$, with vertex operator of the form $V_{4;s}$, and vertices $v_{s'}$, with vertex operator
of the form $V_{4;s'}$, with $s'\neq s$. 
A face $f$ can pass through a vertex $v$ a number of times denoted $\alpha=0,1,2$. We 
symbolically write this as $v^{\alpha} \in f$ and then define a new type of incidence matrix by 
\beq\label{incivf}
\epsilon_{v_sf_{s'}} = \left\{ \begin{array}{cc} \alpha ,& \;\;\text{if } s=s' \text{ and if }\;\;v_s^{\alpha} \in f_{s} ,\\
0,& \;\;\text{otherwise.} \end{array}
\right. 
\eeq
For any color $s$, we write the contribution in the amplitude,
at fixed vertex $v_{s}$, as 
$\lambda(1+ \tilde\eta \sum_{f_{s'}} \epsilon_{v_s,f_{s'}} p_{f_{s'}})$, with  $\tilde\eta=\eta/\lambda$. A noteworthy fact must be reported: calculating bounds on amplitudes in usual/local quantum field theory and
from the simple graph theory perspective, the incidence matrix $\epsilon^{(1)}_{vl}$ encoding the incidence between lines and vertices is of major importance. 
In particular tensorial models \cite{dine1,Carrozza:2013mna}, there is 
another matrix $\epsilon^{(2)}_{lf}$ which encodes the incidence between lines 
and faces and proves also to be useful. These two objects, in colored tensor models, are somehow natural because they are related with the existence of a homological structure on  colored graphs \cite{coloured}.  In the present context, we discover a new matrix $\epsilon_{vf}$ which records the incidence vertices-faces. This is certainly a peculiarity of the present class of models which gives a weight to the vertices. 
On may naturally wonder if the product of the incidence matrices $\epsilon^{(1)}_{vl}$
by $\epsilon^{(2)}_{lf}$ does not generate $\epsilon_{vf}$. The answer to this
question is: ``no, in general,'' $\sum_{l}\epsilon^{(1)}_{vl} \epsilon^{(2)}_{lf} \ne \epsilon_{vf}$.  

Let us introduce the set $\cV_s$  of vertices with vertex 
kernel $V_{4;s}$. Then, $\cV = \sqcup_{s =1}^d\cV_{s}$ (disjoint union). 
Using the Schwinger parametric form of the propagator kernel as
\beq
\tilde C(\{p_s\}) = \int_0^{\infty} d\alpha \; e^{-\alpha(\sum_s p_s^2 + \mu)}\,, 
\eeq
we put  the amplitude \eqref{ampli} in the form 
\bea\label{amplf}
&&
A_{\cG} = \kappa(\lambda)\sum_{p_{f_s}} 
\int \Big[\prod_{l\in \cL} d\alpha_l\, e^{-\alpha_l \mu} \Big]\; 
\prod_{f_s\in \cF_{\ext}} e^{-(\sum_{l\in f_s} \alpha_l) (p^{\ext}_{f_s})^{2}}
\prod_{f_s\in \cF_{\inter}} e^{-(\sum_{l\in f_s} \alpha_l) p_{f_s}^{2}} 
\prod_{s=1}^{d}\prod_{v_s \in \cV_s} [1 + \tilde\eta (\epsilon\,\tilde{p}^{\,2a})_{v_s} ]\,, \cr\cr\cr
&&
(\epsilon\,\tilde{p}^{\,2a})_{v_s} := 
\sum_{f_{s'}} \epsilon_{v_s,f_{s'}}(\tilde{p}_{f_{s'}})^{2a}\,, 
\eea
where $\kappa(\lambda)$ includes symmetry factors
and coupling constants, $p^{\ext}_{f_s}$ are external momenta which are not summed
and, in the last line, $\tilde{p}_{f_s}$ refers to an internal or external momentum. 
Summing over arbitrary high momenta might produce divergent amplitudes \eqref{amplf},
hence the need of a renormalization analysis. 
Our next goal will be not to perform this analysis  but to initiate
that program by providing a power counting theorem for any amplitude. 
The particular
scheme that we use for working out such a theorem 
is the so-called multi-scale analysis \cite{Rivasseau:1991ub}. 
So far, it has been proved enough powerful to address any tensorial 
model at the perturbative level. We will see that, in the present situation as well, the
multi-scale analysis allows to reach a power counting theorem. 

\

\noindent{\bf Multiscale analysis.}
We start the multiscale analysis by introducing 
a slice decomposition of the propagator in the parameter
$M>1$, and determine bounds on each sliced propagator:
\bea\label{bounds}
&&
\tilde C(\{p_s\}) = \int_0^{\infty} d\alpha\; e^{-\alpha(\sum_s p_s^2 + \mu)}
 = \sum_{i=0}^\infty C_i  (\{p_s\})\,, \cr\cr
&&
C_i  (\{p_s\}) = \int_{M^{-2(i+1)}}^{M^{-2i}} d\alpha\;
e^{-\alpha(\sum_s p_s^2 + \mu)} 
\leq K' M^{-2i}  
e^{- M^{-2i}(\sum_s p_s^2 + \mu)}
\leq K M^{-2i}  \;
e^{- \delta M^{-i}(\sum_s |p_s| + \mu)} \,,\cr\cr
&&
C_0 (\{p_s\}) = \int_{1}^{\infty} d\alpha\;
e^{-\alpha(\sum_s p_s^2 + \mu)} \leq K\,,
\eea
for some constants $K$, $K'$ and $\delta$. 
As in the standard field theory case, high values of $i$ select high 
momenta of order $M^{i}$ and this will be called UV.
It is immediate that this regime also coincides with 
small distances on $U(1)$. Meanwhile, small momenta are selected
by the slice $i=0$, and corresponds to the IR. 
We introduce a cut-off $\Lambda$ and the cut-offed
propagator expresses as $C^{\Lambda}=\sum_{i=0}^{\Lambda} C_i$.  

Beginning with the analysis, the next developments  follow  the same steps
as detailed in \cite{rankd} but extra features arise
from the vertex weights and  must be discussed. We slice all propagators in \eqref{ampli} and write, using the bounds \eqref{bounds}, 
\bea
&&
A_{\cG} = \sum_{\bmu} A_{\cG;\bmu}\,, \qquad 
A_{\cG;\bmu} =\sum_{p_{v;s}} 
\prod_{l \in \cL } C_{i_l} (\{p_{v(l)}\}; \{p'_{v'(l)}\}) 
\prod_{v \in \cV} (-V_{4;v} (\{p_{v;s}\})) \,, \cr\cr
&&
|A_{\cG;\bmu}|\leq  \kappa(\lambda) 
K^L
K_1^V  \; K_2^{F_{\ext}} \prod_{l \in \cL} M^{-2 i_l} \, 
\sum_{p_{f_s}} 
\prod_{f_s\in \cF_{\inter}} e^{-\delta(\sum_{l\in f_s} M^{-i_l}) |p_{f_s}|} 
\prod_{s=1}^{d}\prod_{v_s \in \cV_s} [1 + \tilde\eta(\epsilon \,\tilde{p}^{\,2a})_{v_s}]\,,
\label{amplinit}
\eea
where $\bmu=\{i_l\}_{l\in \cL}$ is a multi-index, called momentum assignment, which collects the  propagator indices $i_l \in [0,\Lambda]$,
$K_{1,2}$ are constants. The sum over the momentum assignments 
will be performed only after renormalization according to 
a standard procedure \cite{Rivasseau:1991ub}. The object of interest
is $A_{\cG;\bmu }$ and seeking an optimal bound for that amplitude is our next goal. Mainly, the sum over internal momenta must be performed in 
a way to bring the less possible divergences, i.e. positive powers of
$M^{i}$. This can be done in a way compatible with the Gallavotti-Nicol\`o
tree of quasi-local subgraphs \cite{Galla}.

Let us define the quasi-local or ``dangerous'' subgraphs which are, by essence, intimately related with a notion of locality of the theory. Consider a 
graph $\cG$, with set $\cL$  of lines and set $\cF_{\inter}$ of internal faces. 
Let $i$ be a fixed slice index and define $\cG^i$ the subgraph of
$\cG$ built with lines with indices such that $\forall \ell \in \cL(\cG^i)\cap\cL$, $i_\ell \geq i$. In the case that $\cG^i$ has
several connected components, we note them $G^i_{k}$. We
call $\{G^i_k\}_{(i,k)}$ the set of quasi-local subgraphs. 
Let $g$ be a subgraph of $\cG$ and call $\cL(g)$ and $\cL_{\ext}(g)$, the sets of internal and external lines of $g$, respectively. Given a momentum
assignment $\bmu$ of $\cG$, define $i_{g}(\bmu)=\inf_{\ell \in \cL(g)} i_\ell$ and $e_{g}(\bmu)=\sup_{\ell \in \cL_{\ext}(g)}i_\ell$, then 
 $g$ subgraph of $\cG$ is quasi-local if and only if the following criterion
is satisfied: $i_{g}(\bmu)>e_{g}(\bmu)$. 

There are well-known facts about the set of quasi-local subgraphs $\{G^i_k\}$: it is partially ordered
under inclusion and forms an abstract tree called the Gallavotti-Nicol\`o
(GN) tree (see Figure 12 in \cite{rankd} for an example given in the tensorial setting). 
We want to perform the sums over internal momenta $p_{f_s}$ in \eqref{amplinit}
in a way compatible with the GN tree, that is, such that the result
can be uniquely expressed  in terms of the graphs $G^i_k$.

Performing the sum over internal momenta $p_{f_s}$,
we will need the particular index $i_f = \min_{l\in f} i_l$. 
This index corresponds to a line $l_f$, namely $i_{l_f}=i_f$. 
The following result which is also useful, can be simply obtained:
\begin{proposition} \label{propsum} Given $n\in \N^*$ and a constant $B\in \R^*$, 
\bea
\sum_{p=1}^{\infty} p^{n} e^{-B p} = c \; B^{-(n+1)} (1+ O(B^{(n+1)})) \,, 
\eea
where $c$ is a constant depending on $n$. 
\end{proposition}
From the above, we can recognize that, at leading order, the discrete sum evaluates like an integral.

We are now in position to  find an optimal bound for the amplitude. 
Because  all external momenta  are chosen such that $p^{\ext}_{f_s} \ll p_{f_{s'}}$, 
for all $s,s'$, and  there is no sum over these external momenta, one
 proves that the following bound holds
\beq
|A_{\cG;\bmu}|\leq  \kappa(\lambda)
K^L K_1^V  \; K_2^{F_{\ext}}\prod_{l \in \cL} M^{-2 i_l} \, 
\sum_{p_{f_s}} 
\prod_{f_s\in \cF_{\inter}} e^{-\delta(\sum_{l\in f_s} M^{-i_l}) |p_{f_s}|} 
\prod_{s=1}^{d}\prod_{v_s \in \cV_s} [1 + K_3\,\tilde\eta(\epsilon \,p^{\,2a})_{v_s}]
\label{ineq}
\eeq
where $K_3$ is another constant, and the matrix $\epsilon $ \eqref{amplf} is now reduced
 to internal faces (we keep however the same notation $\epsilon$).

The sum over internal momenta $p_{f_s}$ must be performed in an optimal way, namely,
in a way bringing the less possible divergences while the inequality \eqref{ineq} must remain a correct approximation. Each sum brings, according
to Proposition \ref{propsum}, a ``bad'' factor of $M^{+i}$. 
The optimal way to evaluate the sum is made in two steps: given a $f$ (forgetting a moment 
the subscript $s$), among the lines $l \in f$, we will simply use the line $l_f$ with $i_{l_{f}}=\min_{l\in f} i_l=i_{f}$, which will generate the lowest factor $M^{i_{f}}$,
such that the sum over $(p_{f}^{2a})^{\alpha_{p_f}} e^{-\delta M^{-i_f} |p_{f}|}$, with 
$\alpha_{p_f}$ an integer,  is the lowest possible. 
Then, there is another difficulty which is to optimize the products of the vertex kernels. As we are searching an upper 
bound for the amplitude and the product generates a sum of positive terms, then we 
must target, in each factor of the product of the vertex kernels, the term $p_f$ generating  after summation a product of $M^{i_f(2a \alpha +1)}$ with the largest possible power.  Thus, there is a monomial generated by 
the product over vertex kernels which yields an upper bound 
of an optimal kind for $A_{\cG;\bmu}$. 
To identify this monomial, we must investigate the combinatorics of the
$\epsilon$ matrix. 

Let us define the following optimization procedure using the 
matrix $\hat\epsilon_{v_sf_{s'}}$ obtained from $\epsilon_{v_sf_{s'}}$ \eqref{incivf} by simply putting all $\alpha=1$. Thus $\hat\epsilon$ simply notes if a face is incident to a given vertex
(the information about how many times this happens contribute to an overall 
constant factor in the bound amplitude \eqref{ineq}). 
 First, for convenience, we organize the incidence matrix $\hat\epsilon$ by color blocks: we  first list  all vertices $v_s$ of a given color as columns and list all 
faces $f_s$ of the same color in row blocks. Next, organizing the line indices for fixed color $s$, we list the faces $f_{s;k}$ from the highest index $i_{f_{s;k}}$ to the lowest; if several
faces have the same momentum index, we list them in an arbitrary way. Start with the face $f_{s;1}$,
and count $\varrho_{f_{s;1}}=\sum_{l}\hat\epsilon_{v_{s,l}f_{s.1}}$, i.e. the number of vertices $v_{s;l}$ such that $\hat\epsilon_{v_{s;l}f_{s;1}}=1$. Then, delete all these columns and the line $f_{s;1}$ and define a new reduced matrix that we denote again, for simplicity, $\hat\epsilon$. 
Pass to the next line $f_{s;2}$ and proceed in the same way with the reduced matrix. 
If there are no more vertices or the matrix trivializes, we define
\beq\label{varro}
\varrho(\cG) = \sum_{s} \sum_{f_{s;k}} \varrho_{f_{s;k}}\,.
\eeq
An illustration of this procedure is given in Appendix \ref{app:exa}. 
At an intermediate step labeled by $f_{s}$, $\varrho(\cG)$ increases its value by the number of times that 
this face $f_{s}$ passes through remaining vertices $v_{s}$
where it still possesses the dominant index $i_{f_s}$. It is immediate
that $\varrho(\cG)$ is bounded from above by the number
of vertices of $\cG$, thus 
\beq
\varrho(\cG)  \leq V(\cG)\,.
\eeq
In some case, $\varrho(\cG)$ coincides with the rank of $\hat\epsilon$ but, of course, it is not in general.

The above procedure leads us to 
an optimal bound and we can observe how this model has an interesting feature:
it combines both an optimization which lowers the value of the indices
of faces while, due to the vertex $p^{2a}_s \phi^4_s$, enhances the contribution
by taking, in an independent way, the largest values among the $i_f$'s. 
We write a new bound
\beq
|A_{\cG;\bmu}|\leq  \kappa_1
 K^L  K_1^V  \; K_2^{F_{\ext}}\prod_{l \in \cL} M^{-2 i_l} \, 
\sum_{p_{f_s}} 
\prod_{f_s\in \cF_{\inter}} e^{-\delta M^{-i_{f_s}} |p_{f_s}|} 
 \prod_{s'=1}^{d}\prod_{f_{s'} } p^{\,2a \varrho_{f_{s'}}}_{f_{s'}}\,, 
\eeq
where $\kappa_1$ incorporates coupling constants $\lambda$ 
and $\eta$. Performing the sum over internal momenta, one
gets
\beq
|A_{\cG;\bmu}|\leq  \kappa_2
 \prod_{l \in \cL} M^{-2 i_l} \, 
\prod_{f_s\in \cF_{\inter}} M^{i_{f_s}(2a\varrho_{f_{s}}+1)} \,, 
\eeq
where $\kappa_2$ is another constant depending on the graph and
including $K, K_1, K_2$ and $\kappa_1$ and new constants
coming from the summation over internal momenta.

\begin{figure}[h]
 \centering
     \begin{minipage}{.7\textwidth}
\includegraphics[angle=0, width=6cm, height=1.5cm]{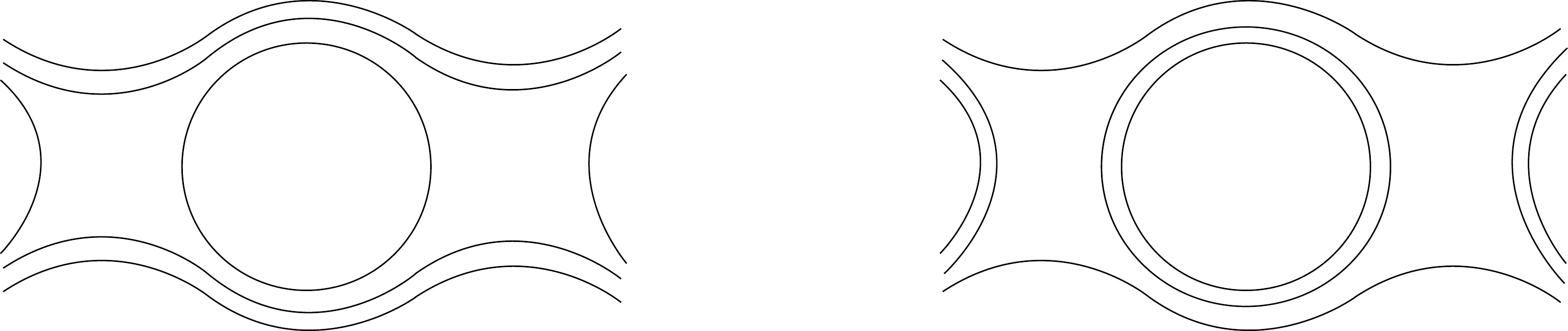} 
\caption{ {\small Two rank 3 4-point graphs: $\cG_1$ is a not a melon
and $\cG_2$ is. }} 
\label{fig:mels}
\end{minipage}
\put(-280,0){ $\cG_1$}
\put(-88,0){$\cG_2$}
\end{figure}

We now use the expansion in quasi-local subgraphs $G^{i}_k$ to write 
the above bound in the form compatible with the Gallavotti-Nicol\`o tree:
\bea
|A_{\cG;\bmu}| &\leq&  \kappa_2
\Big[ \prod_{l \in \cL}\prod_{i=1}^{i_l} M^{-2}\Big] \, 
\prod_{f_s\in \cF_{\inter}} \Big[\Big(\prod_{i=1}^{i_{f_s}} M \Big)
\Big(\prod_{i=1}^{i_{f_s}} M^{2a \varrho_{f_{s}}} \Big)
\Big]\\
&\leq& 
 \kappa_2
\Big[ \prod_{l \in \cL}\prod_{(i,k)/\, l\in \cL(G^i_k)} M^{-2}\Big]  \, 
\prod_{f_s\in \cF_{\inter}} 
\Big[\Big(\prod_{(i,k)/\, l \in \cL(G^{i_{f_s}}_k)}  M \Big)
\Big(\prod_{(i,k)/\, l\in \cL(G^{i_{f_s}})} M^{2a \varrho_{f_{s}}} \Big)
\Big].\nonumber
\eea
The product $ \prod_{l \in \cL}\prod_{(i,k)/\, l\in \cL(G^i_k)} M^{-2}$ can 
be recast in a standard way \cite{Rivasseau:1991ub}, as $\prod_{(i,k)} M^{-2 L(G^i_k)}$. 
The second product $\prod_{f_s\in \cF_{\inter}} 
\Big(\prod_{(i,k)/\, l \in \cL(G^{i_{f_s}}_k)}  M \Big)$ has been studied
in previous works \cite{bgr,rankd} as well. We re-express it as 
\bea
\prod_{f_s\in \cF_{\inter}} 
\prod_{(i,k)/\, l \in \cL(G^{i_{f_s}}_k)}  M 
 = \prod_{ f_s\in \cF_{\inter} } \prod_{(i,k)/\, l_{f_s} \in \cL(G^{i}_k)}  M 
 = \prod_{(i,k)} \prod_{ f_s\in \cF_{\inter} \cap G^{i}_k} M
 =  \prod_{(i,k)} M^{F_{\inter}(G^i_k)} \,,
\eea
where we use the fact that the face $f$ becomes closed in the graph  $G^i_k$
if the line $l_f \in \cL(G^i_k)$. The last product, namely, the one involving the new ingredient $\varrho_{f_s}$, introduces a new feature for the present model. We address it in the following way:
\bea
\prod_{ f_s\in \cF_{\inter} } 
\prod_{ (i,k)/\, l_{f_s}\in G^{i}_{k}  }M^{2a \varrho_{f_s}} 
= \prod_{(i,k)} \prod_{ f_s\in \cF_{\inter}\cap G^i_k }
M^{2a \varrho_{f_s} } = \prod_{(i,k)} M^{2a \varrho(G^i_k)}\,,
\eea
where $\varrho(\cdot)$ has been defined in \eqref{varro}. 
We reach the following statement:
\begin{theorem}[Power counting]\label{theo:pc}
Let $A_{\cG;\bmu}$ be the amplitude associated with the graph $\cG$ of the  $p^{2a}\phi^4_d$-model in the multi-scale index $\bmu$, then there exists a constant $\kappa$ depending
on the graph such that 
\beq
|A_{\cG;\bmu}| \leq \kappa \prod_{(i,k)\in \N^2} M^{\omega_{\rmd}(G^i_k)} \,, 
\eeq
where $G^i_k$ are quasi-local subgraphs and  
\beq
\omega_{\rmd}(G^i_k) = -2 L(G^i_k) + F_{\inter}(G^i_k) +2a \varrho(G^i_k)\,. 
\eeq
\end{theorem}
Setting $a\to 0$ brings us back to the degree of divergence of usual
tensorial models \cite{bgr,rankd}.  The term $2a \varrho(G^i_k)$ enhances, as predicted and {\it a priori}, the divergence degree of any graph. In particular, it allows non-melonic graphs to diverge as well. Indeed, consider the non-melonic 4-point graph $\cG_1$ of Figure \ref{fig:mels}. Such a graph has a (superficial) degree of divergence:
\bea
\omega_{\rmd}(\cG_1) =  -2\times 2 + 1+2a \times 2 = 4a -3
\eea
which is strictly positive, and so possesses a divergent amplitude, whenever $a> \frac34$.
Evaluating the 4-point melonic graph $\cG_2$ in the same figure, one finds $\omega_{\rmd}(\cG_2)=-2\times 2 +2 =-2<0$ which implies a convergent amplitude. 
By this calculation, at large momenta, we simply realize that non-melonic diagrams can dominate melonic ones. 
This also shows that the type of graphs that need to be renormalized, in
this framework, is radically different from any type known in previous tensorial
models. In top of melonic graphs, we might have non-melonic graphs which
might contribute to the flow of the coupling constants. However, the issue of 
renormalization is very subtle in general, and surely more, in the present context.
For a given graph $\cG$, the quantity $\varrho(\cG)$ must be fully analyzed before addressing the second stage for a renormalization procedure which should consist in identifying a locality principle for this class of models. This is left to a subsequent work.
Nevertheless, even at this stage, from the above power counting theorem, we can extract additional information on the model which could enlighten its renormalizability property.
For instance, we can discuss now how to fix the rank $d$ and parameter $a$ in a way 
which might lead to renormalizable models given a maximal valence of the interaction $\phi^{k_{\max}=4}$. The list the divergent graphs and their boundary data
coming from these models will determine precisely the locality principle mentioned above. 
To proceed with this locality and renormalizability study, another important fact must be investigated:  
is there a hidden competition between $F_{\inter}$ and $\varrho$, such that $\varrho$ cannot be large if $F_{\inter}$ is, and vice-versa? Our next analysis does not take any consideration 
of the kind and thus,  it is possible that it might be  improved
(for e.g. turning sufficient conditions to necessary and sufficient 
ones).

Let $\cG$ be a graph, then, as aforementioned $\varrho(\cG)\leq V(\cG)$, and so we do have (from now on, we omit the dependency of the graph $\cG$ in the basic quantities
$V=V(\cG)$, $L=L(\cG)$, etc...)
\beq
\omega_{\rmd}(\cG) \leq -2 L + F_{\inter} + 2a V\,. 
\eeq
and this bound can be saturated. 

Let us note  
that the number of internal faces of a connected graph $\cG$, in any rank $d\geq 3$ tensorial model, is given by  \cite{dine1}:
\beq
\label{faceinter}
F_{\inter} = -\frac{2}{(d-1)!}(\omega(\cexG) - \omega(\bG)) - (C_\bG -1)
- \frac{d-1}{2} N_{\ext} + d-1 - \frac{d-1}{4}(4-2n)\cdot V\,,
\eeq
where $\cexG$ is called the colored extension of $\cG$ (Definition 1$i$ in \cite{bgr}),
$\bG$ defines the boundary of $\cG$ (Definition 1$iv$ in the same reference above), with number $C_\bG$  of connected components, $V_k$ its number of vertices of coordination $k$, $V= \sum_{k} V_k$ its total number of vertices,  $n \cdot V = \sum_{k} k V_k$ its number of half lines exiting from vertices, $N_{\ext}$ its number of external legs. 
 The number $\omega(\cexG)=\sum_{J}g_{\tJ}$ is called the degree of $\cexG$, 
$\tJ$ is the ``pinched'' jacket associated with $J$, a jacket of $\cexG$ (Definition 1$iii$), 
$\omega(\bG)=\sum_{\bJ}g_{\bJ}$ is the degree of $\bG$ (Definition 1$v$). 
The interested reader can have a proof of the formula \eqref{faceinter} after Proposition 3.7 in \cite{dine1}.  Using now the combinatorial formula
\beq
-2L = -(n \cdot V - N_{\ext})\,, 
\eeq
we obtain, setting $d^-=d-1$,
\bea
&&
\omega_{\rmd}(\cG) =
-\frac{2}{(d^-)!}(\omega(\cexG) - \omega(\bG)) - (C_\bG -1)
- \frac{d^-}{2} N_{\ext} + d^- - \frac{d^-}{4}(4-2n)\cdot V 
- (n \cdot V - N_{\ext}) + 2a \varrho
\cr\cr
&&
=  -\frac{2}{(d^-)!}(\omega(\cexG) - \omega(\bG)) - (C_\bG -1)
- \frac12\big(   (d^--2) N_{\ext} - 2d^-  \big)
-\frac12 \big[ \big(  2d^-     + ( 2 - d^-) n \big)\cdot V - 4 a\varrho\big].
\label{centr}
\eea
It has been proved in \cite{add}, that either $\omega(\cexG) -\omega(\bG)=0$ or it satisfies
the bound  
\beq\label{boundOm}
-\frac{2}{(d^-)!}(\omega(\cexG) -\omega(\bG))\leq -(d^--1)  \,. 
\eeq

Thus, using \eqref{boundOm}, we have the following bounds 

\noindent$\bullet$ $\omega(\cexG)-\omega(\bG)=0$ and 
\beq\label{bou1}
\omega_{\rmd}(\cG) \leq   - (C_\bG -1)
- \frac12\big(   (d^--2) N_{\ext} - 2d^-  \big)
-\frac12 \big[ \big(  2d^-     + ( 2 - d^-) n \big)\cdot V -4a\varrho\big] \,;
\eeq
\noindent$\bullet$ $\omega(\cexG)-\omega(\bG)>0$  and 
\beq\label{bou2}
\omega_{\rmd}(\cG) \leq  -(d^- -1) - (C_\bG -1)
- \frac12\big(   (d^--2) N_{\ext} - 2d^-  \big)
-\frac12 \big[ \big(  2d^-     + ( 2 - d^-) n \big)\cdot V - 4 a\varrho \big]\, .
\eeq

\

\noindent{\bf On potentially renormalizable models.}
We investigate the particular cases of 
ranks $d=3$ and $4$ and restrict to $k_{\max}=4$ as the maximal valence of the 
theory vertices. Hence, $V= V_2 + V_4$ and $n \cdot V = 2V_2 + 4 V_4$. 
We now aim at (1) finding sufficient conditions
such that only graphs with at most four external fields might
diverge and at (2) showing that, for some values of $a$, there are non-melonic graphs which might be involved in the renormalization 
analysis. The number $V_2$ of mass vertices does not add much to the discussion
in the following, we will simply neglect it.

\begin{enumerate}
\item[(i)] In rank 3 ({\bf $\phi^{4}_{d=3}$-model}), $d^- =2$, 
the relation \eqref{centr}  exhibits a crucial fact: the divergence degree does not depend
on the number of external legs of the graph. The bounds \eqref{bou1}
and \eqref{bou2} take the form

\noindent$\bullet$ $\omega(\cexG)-\omega(\bG)=0$ and 
\beq\label{bou11}
\omega_{\rmd}(\cG) \leq   - (C_\bG -1) -2( V_4-1-a\varrho) \,;
\eeq
\noindent$\bullet$ $\omega(\cexG)-\omega(\bG)>0$  and 
\beq\label{bou21}
\omega_{\rmd}(\cG) \leq - (C_\bG -1)
- [2(V_4 -  a\varrho) -1]\,.
\eeq

We start by the second inequality \eqref{bou21} which is 
quite coercive. Whenever $2(V_4-a\varrho) \leq 1$, we might have divergent graphs. 
As we can build  graphs with number of external fields higher than 4 which  diverge and satisfy that condition (see, for e.g., the 6-point graph in Figure \ref{fig:phi6} which is linearly divergent with $C_{\bG}=1$, $\varrho =V_4$ and fixing $a=1$,
the external data of which do not correspond to a term initially present in our action \eqref{eq:3dkin} and \eqref{eq:3dinter}\footnote{Note that 
one can easily generalize this example to an arbitrary number of legs $N_{\ext}=2n\geq 2$, $V_{4}=n$, the divergence occurs when $1+2a n >2 \times n$, i.e. when $a>1-\frac{1}{2n}$. The case $n=1=V_4$ is discussed in the following as the non-melonic tadpole, Figure \ref{fig:tad}; the case $n=2$ was already discussed in Figure \ref{fig:mels}. }), thus,  without any further assumptions, this is a signal of a nonrenormalizable $\phi^4$-model. 
We must therefore find conditions which can lead us to a better control on the number
of external legs. 

\begin{figure}[h]
 \centering
     \begin{minipage}{.7\textwidth}
\includegraphics[angle=0, width=3.5cm, height=2cm]{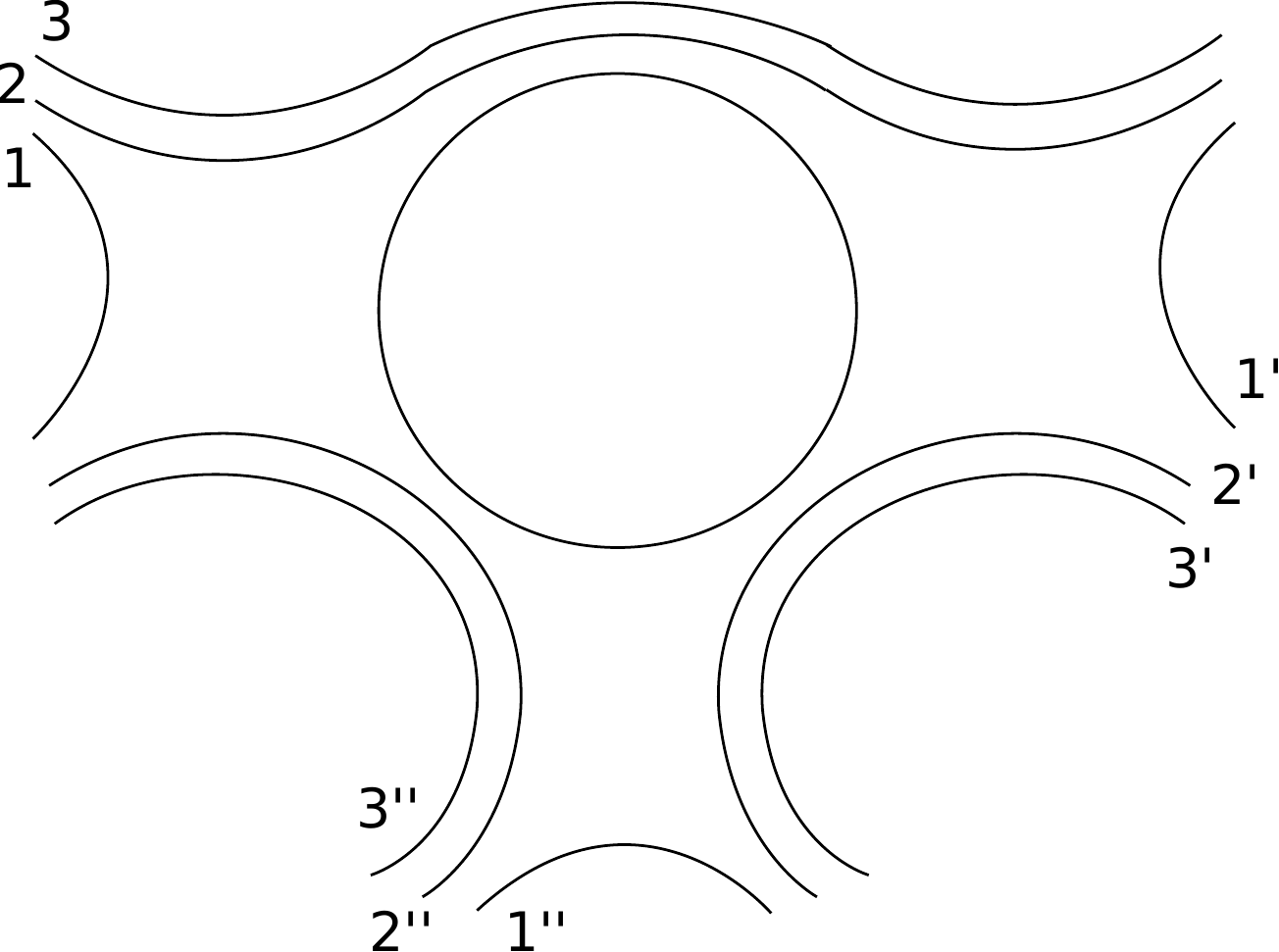} 
\caption{ {\small  A rank 3 $\phi^6$-graph.}} 
\label{fig:phi6}
\end{minipage}
\end{figure}

$\sim$ (i1) The case $V_4=1$ is very particular: these are tadpole graphs with  one propagator (to avoid triviality), 
$C_{\bG}=1$, $N_{\ext}=2$ and $\omega_{\rmd} (\cG) \leq -1+2a\varrho$. 
Since $\varrho\leq V_4=1$, $\varrho$ can only be $0$ or $1$.
The case leading to divergence is given by $\omega_{\rmd} (\cG) =-1+2a$. This is a non-melonic tadpole (see Figure \ref{fig:tad})
with $\omega(\cG_{\ext})-\omega(\bG)=1$. This contribution is
divergent if $a \geq 1/2$, and converges otherwise.  Any divergence must be re-absorbed by the mass or wave function renormalization.

\begin{figure}[h]
 \centering
     \begin{minipage}{.7\textwidth}
\includegraphics[angle=0, width=1.4cm, height=1.4cm]{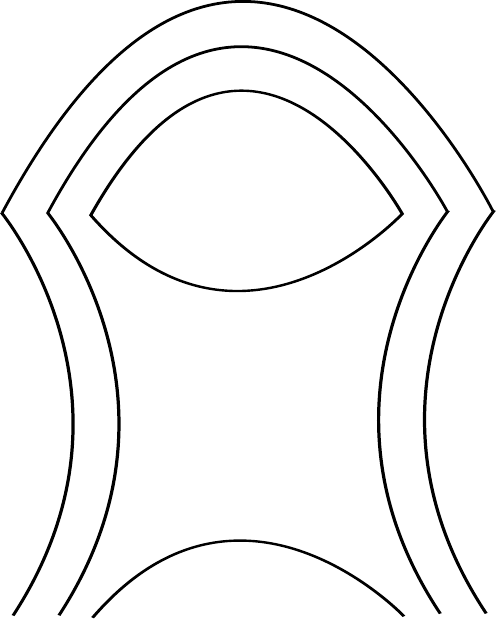} 
\caption{ {\small  Rank 3 non-melonic tadpole.}} 
\label{fig:tad}
\end{minipage}
\end{figure}

$\sim$ (i2)
Now, consider $V_4= 2$ and a number of external legs $2 \leq N_{\ext} \leq 4$ (noting that $N_{\ext}=6$ defines a graph without loops or disconnected). 
We get $\omega_{\rmd} (\cG) \leq -3+2a\varrho\leq -3 + 4a$ which 
can be positive ($a \geq \frac34$) or negative. 
Any divergences must be tackled by mass and wave function, if $N_{\ext}=2$,  and $\phi^4$-coupling,  if $N_{\ext}=4$.

$\sim$ (i3) We finally consider $V_4 \geq 3$ and $N_{\ext}\geq 2$. 
Then a solution of the renormalizability problem is to enforce a sufficient condition making all these amplitudes convergent. And such a condition is given by $2(V_4-a\varrho ) >1$, $\forall V_4\geq 3$. In that situation, one finds $ 0< a < \frac56 \leq  1 -\frac{1}{2V_4}$.

The first relation \eqref{bou11} dealing with only melonic graphs is now discussed. From the above analysis, we restrict the investigation to the interval $a\in (0,\frac56)$.  Before undertaking any case-by-case
analysis depending on the number of vertices, one must
observe that if $\varrho =0$, $\omega_{\rmd}(\cG)$ can
be at most 0, and this occurs if $C_\bG =1$, since $V_4 \geq 1$. 
 The possible divergent terms will renormalize 
the mass (these  are log-divergent terms and the expansion of the amplitude generates subleading terms which are  
convergent). As noticed, we do not have any constraint on $a$. 
Now concentrating $\varrho>0$,

$\sim$ (ii1) $V_4=1$, there is a single possibility to construct
a melonic tadpole and for this case $\varrho=0$. 

$\sim$ (ii2) $V_4= 2$, $N_{\ext}=2$, there is
a finite number of ways to construct melonic graphs such that $\varrho>0$, by mixing $V_{4;s}$ and $V_{4;s'}$, $s\neq s'$.   
For such graphs, $\varrho=1$, $C_\bG =1$, $\omega_{\rmd}(\cG)\leq 
-2(2-1-a) = -2(1-a)$ which cannot diverge if $a<1$. 
For $N_{\ext}=4$, there is no such occurrence. 

$\sim$ (ii3) $V_4\geq 3$, a condition making all amplitudes
convergent is $V_4-1-a\varrho>0$. This is satisfied if $a<1-\frac{1}{V_4}$. Thus $0<a<\frac{2}{3}$ guarantees the convergence 
of all such amplitudes, and this hints at super-renormalizability. 
In the case $\frac{2}{3}\leq a <\frac56$, one can 
show that, for $V_4 \leq 6$, the amplitudes might diverge
and whenever $V_4>6$, we necessarily have a convergent
amplitude. As a result, the number of divergent configurations is finite and this is a strong signal of super-renormalizability. The sole
problem is that, in this case, one must check 
if there is no $(N>4)$-graph which  diverge. If yes then,
it means that $\phi^4$-model is not renormalizable in this truncation.

As a result, the above analysis suggests  the following table 
(below tadpole graphs are with $V_4=1,2$)
\bea
0 < a< \frac12 \,, && \quad  p^{2a}\phi_{d=3}^4\; \text{ is super-renormalizable with melonic divergent tadpole graphs; }
\cr\cr
\frac12\leq a < \frac23 \,, && \quad  p^{2a}\phi_{d=3}^4\; \text{ is super-renormalizable with melonic and non-melonic divergent tadpole graphs; }
\cr\cr
\frac23\leq a < \frac56 \,, && \quad  \text{Inconclusive: melonic and non-melonic tadpole
diverge; non-melonic 4pt-graphs diverge} \cr\cr
&&\quad \text{ if $a\geq \frac34$; melonic
graphs with $\varrho>0$ and $V_4\leq 6$ might diverge;}
\cr\cr
a \geq \frac56 \,, && \quad p^{2a}\phi_{d=3}^4\; \text{ is  non-renormalizable. }
\eea

\item[(ii)] In rank 4 ({\bf $\phi^{4}_{d=4}$-model}), $d^- =3$, the situation is a little 
more involved but, still, we can address it in a similar way as above. 
We have:

\noindent$\bullet$ $\omega(\cexG)-\omega(\bG)=0$ and 
\beq\label{bou13}
\omega_{\rmd}(\cG) \leq   - (C_\bG -1)
- \frac12\big(   N_{\ext} - 6 \big)
- (V_4 -2a \varrho)\,;
\eeq
\noindent$\bullet$ $\omega(\cexG)-\omega(\bG)>0$ and 
\beq\label{bou23}
\omega_{\rmd}(\cG) \leq   - (C_\bG -1)
- \frac12\big(  N_{\ext} - 6  \big)
-\big( V_4 +2- 2 a\varrho \big)\, .
\eeq
The case $N_{\ext}=6$ is enough particular to be stressed right
away: it seems that a $p^{2a}\phi^6$-model would have, in this rank,
interesting renormalizability properties. Indeed such a model would already have   features similar to the model $\phi^6$ of \cite{bgr} (divergent melonic contributions with $\varrho=0$ from \eqref{bou13}). 
Nevertheless, it will require a greater challenge to work out in
details, because the model will have with two types of
trace invariants $\Tr_{6;1;i}(\phi^6)$ and $\Tr_{6;2;ij}(\phi^6)$ invariants
(with $i$ and $j$ parametrizing color indices). 

To start with, let us consider a graph with $N_{\ext}\geq 6$, then 
we obtain from the previous bounds

\noindent$\bullet$ $\omega(\cexG)-\omega(\bG)=0$ and 
\beq\label{bou131}
\omega_{\rmd}(\cG) \leq   - (C_\bG -1)
- \big(V_4- 2 a\varrho \big) \,;
\eeq
\noindent$\bullet$ $\omega(\cexG)-\omega(\bG)>0$  and 
\beq\label{bou231}
\omega_{\rmd}(\cG) \leq   - (C_\bG -1)
-\big( V_4 +2- 2 a\varrho \big)\, .
\eeq
Noting that $V_4- 2 a\varrho \leq V_4+2- 2 a\varrho$, 
 for having all graphs convergent, 
it is sufficient to require from \eqref{bou131} that $V_4- 2a\varrho \geq V_{4}(1-2a)>0$ implying $a<\frac12$. A rapid checking shows that, indeed, 
for $a\geq \frac12$, there are graphs with higher valency 
which diverge and so the model is non-renormalizable. 
A specific example is given in  Figure \ref{fig:6pt}. Fixing $a=\frac34$, we have a quadratic divergent amplitude which is of the $\phi^6$-form.

\begin{figure}[h]
 \centering
     \begin{minipage}{.7\textwidth}
\includegraphics[angle=0, width=3.5cm, height=2.5cm]{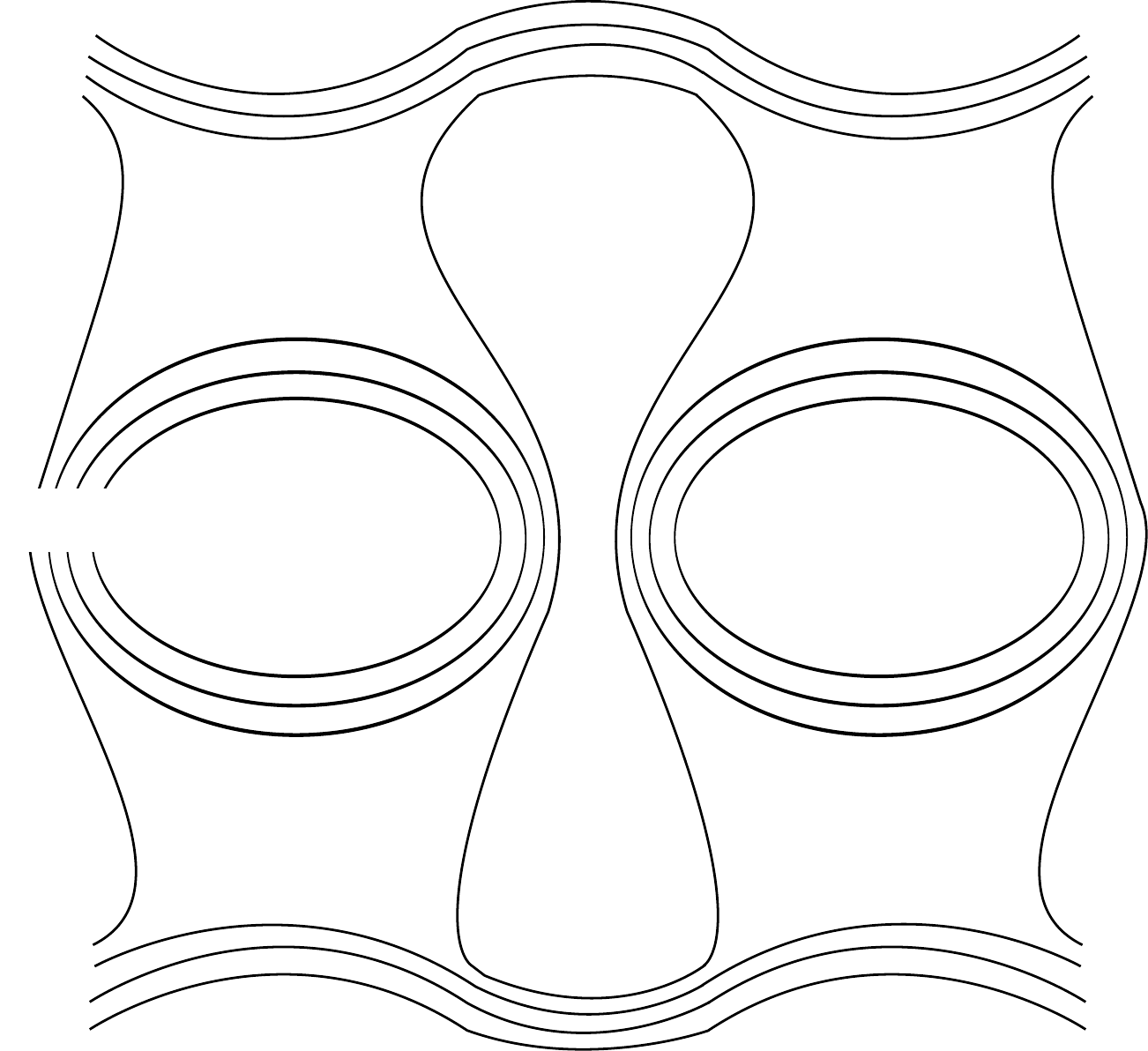} 
\caption{ {\small  Rank 4 divergent 6-point graph.}} 
\label{fig:6pt}
\end{minipage}
\end{figure}

Now, we treat the case of a graph with $N_{\ext}=4$ and $V_4 \geq 2$ 
($V_4<2$ is trivial) with the bounds

\noindent$\bullet$ $\omega(\cexG)-\omega(\bG)=0$ and 
\beq\label{bou133}
\omega_{\rmd}(\cG) \leq   - (C_\bG -1)
- ( V_4 -1 -2a\varrho)\,;
\eeq
\noindent$\bullet$ $\omega(\cexG)-\omega(\bG)>0$ and 
\beq\label{bou233}
\omega_{\rmd}(\cG) \leq   - (C_\bG -1)
-\big( V_4 +1- 2 a\varrho \big)\, .
\eeq
Let us set $\varrho=0$, then it is immediate
that all amplitudes under this condition are convergent:
$1< V_4 -1 < V_4 +1$ and $\omega_{\rmd}(\cG)<0$. We now focus on $\varrho>0$. 
From the study of $N_{\ext} \geq 6$, we must fix
 $0<a < \frac12$. Because $\varrho\leq V_4$, 
$1 < V_4+1 - 2a\varrho$, and then we infer from the relation \eqref{bou233} that all non-melonic graph amplitudes with 
$\varrho>0$ are convergent. 
The relation \eqref{bou133} is now scrutinized at $\varrho>0$
and $0<a < \frac12$. One must observe that,
if $V_4-\varrho=x\geq 1$  then $V_4-1-2a\varrho = V_4 -1 -2a (V_4-x)
= (V_4-1)(1-2a)+2a(x-1)>0$. Therefore, the only 
case which might bring divergence is given by $\varrho=V_4$. 
In the interval $-1<V_4(1 -2a)-1\leq 0$, we could generate divergences 
and this is choosing $a $ such that  $\frac12(1-\frac{1}{V_4}) \leq a < \frac12$, for all $V_4$.
If there exists an infinite family of graphs with arbitrary 
number of vertices $V_4$ such that $a$ satisfies this bound,
this means that $\frac12 \leq a <\frac12$ which is absurd.
So the only way to have divergent graphs is that these belong
to a finite family  with a maximal finite number  $V_4$.  
This suggests super-renormalizability.   
At this stage, we did not have any graph example
such that $\varrho=V_4\geq 2$ which is melonic and divergent. 
What is obviously true instead is that restricting $a <\frac14$ leads directly  to the convergence of any amplitude. 
Thus, at this stage, for $\frac14 \leq a< \frac12$, the model renormalizability behavior is not determined.

Next, the case $N_{\ext}=2$, $V_4 \geq 1$, is now clarified
for $0<a < \frac14$. 

\noindent$\bullet$ $\omega(\cexG)-\omega(\bG)=0$  and 
\beq\label{bou134}
\omega_{\rmd}(\cG) \leq   - (C_\bG -1) - (V_4-2-2a\varrho) \,;
\eeq
\noindent$\bullet$ $\omega(\cexG)-\omega(\bG)>0$ and 
\beq\label{bou234}
\omega_{\rmd}(\cG) \leq   - (C_\bG -1) 
-\big( V_4 - 2 a\varrho \big)\, .
\eeq
Let us inspect the case $\varrho=0$,  \eqref{bou234} and $V_4\geq 1$
mean convergence for amplitudes of non-melonic graphs,  and \eqref{bou134} shows that 2-point melonic graphs can be at most linearly divergent if $V_4\leq 2$ and they are convergent otherwise. They could renormalize the mass and possibly the wave function. 

We focus now on $\varrho >0$. From \eqref{bou234},
$V_4 - 2a \varrho \geq  V_4 (1 -2a) > 0 $, thus all non-melonic
graphs of this kind are finite. Starting by \eqref{bou134}, 
we have 
$\frac{V_4}{2}-2 \leq V_4 - 2 -\frac{\varrho}{2}< V_4 - 2 -2a \varrho<V_4-2 $.
As a result, all graphs with $V_4 >4$ are convergent. 
There is a finite number of configurations and 
the maximal degree of divergence is 2. These graph amplitudes
must renormalize the mass and wave function, at subleading order.

The study of the rank $d=4$ suggests that 
\bea
0 < a < \frac14 \,, && \quad  p^{2a}\phi_{d=4}^4\; \text{ is super-renormalizable with divergent melonic  2-point }\cr\cr
&& \text{  graphs made with $V_4 \leq  4$ vertices; }
\cr\cr
\frac14\leq a < \frac12 \,, && \quad  \text{Inconclusive: melonic  4pt-graphs might diverge;} 
\cr\cr
 a \geq  \frac12 \,, && \quad p^{2a}\phi_{d=4}^4\; \text{ is  non-renormalizable. }
\eea
\end{enumerate}

\section{Conclusion}
\label{ccl}

We have introduced a tensorial field theory of the type $p^{2a}\phi^4$ and studied
its power counting theorem using multi-scale analysis. 
We find a new degree of divergence of the amplitudes generalizing
that of \cite{bgr} and \cite{rankd} by adding a new quantity related to the incidence
matrix of vertices and faces. The main motivation for introducing such a $p^{2a}\phi^4$ model is to inquire other types of continuum limits and phases  different from the branched polymer one, namely, 
the continuum limit of the simplest tensorial models. 
The present model shows that previous  suppressed contributions
(called non-melonic graphs) in ordinary tensorial models become enhanced in the present context. At the field theory level, our present analysis reveals that, indeed, non-melonic
contributions can be of the same degree of divergence than the melonic ones and even, in some case, more relevant. 
This is encouraging for the above program on the continuum limit. 
Concerning renormalizable models, we have strong indications that, for a range of value of $a$, there are
several $\phi^4$-models which might be super-renormalizable although
did not have yet any hint for just-renormalizable models of the $\phi^4$-type. 
In general, the role played by non-melonic contributions is not yet really sensible
in the $\phi^4$-truncation.  
The $\phi^6_{d=4}$ model 
seems to possess relevant properties within the above scheme. 

Finally, let us mention that other exotic choices for models are possible to complete the $p^{2a}\phi^4$-model. For instance, we can introduce a different power of the momenta in the kinetic term $ \bdel_{p_i,p_i'} ( \sum_{i=1}^d p^{2b}_i)$ (as highlighted in \cite{rankd}) which will allow to explore more models in two parameters $(a,b)$ and might be a better approach to find just-renormalizable models. We might also change the dimension
of the group from $U(1)$ to $U(1)^D$ and perhaps consider the group $SU(2)$
in the way of \cite{rankd}. 
This must be fully addressed elsewhere.

\section*{Acknowledgements}
The author thanks Vincent Rivasseau fruitful discussions and for having drawn his attention on this class of models. He is also deeply grateful for sharing with him, in a spirit of friendship and search for truth, insights in mathematical and theoretical physics. 
 
 J.B.G. acknowledges the support of the Alexander von Humboldt Foundation
and the Max-Planck Institute for Gravitational Physics, Albert Einstein Institute. 

\section*{ Appendix}
\label{app}

\appendix

\renewcommand{\theequation}{\Alph{section}.\arabic{equation}}
\setcounter{equation}{0}

\section{Optimization of the momentum sums: a worked out example}
\label{app:exa}

\begin{figure}[h]
 \centering
     \begin{minipage}{.7\textwidth}
\includegraphics[angle=0, width=6cm, height=4cm]{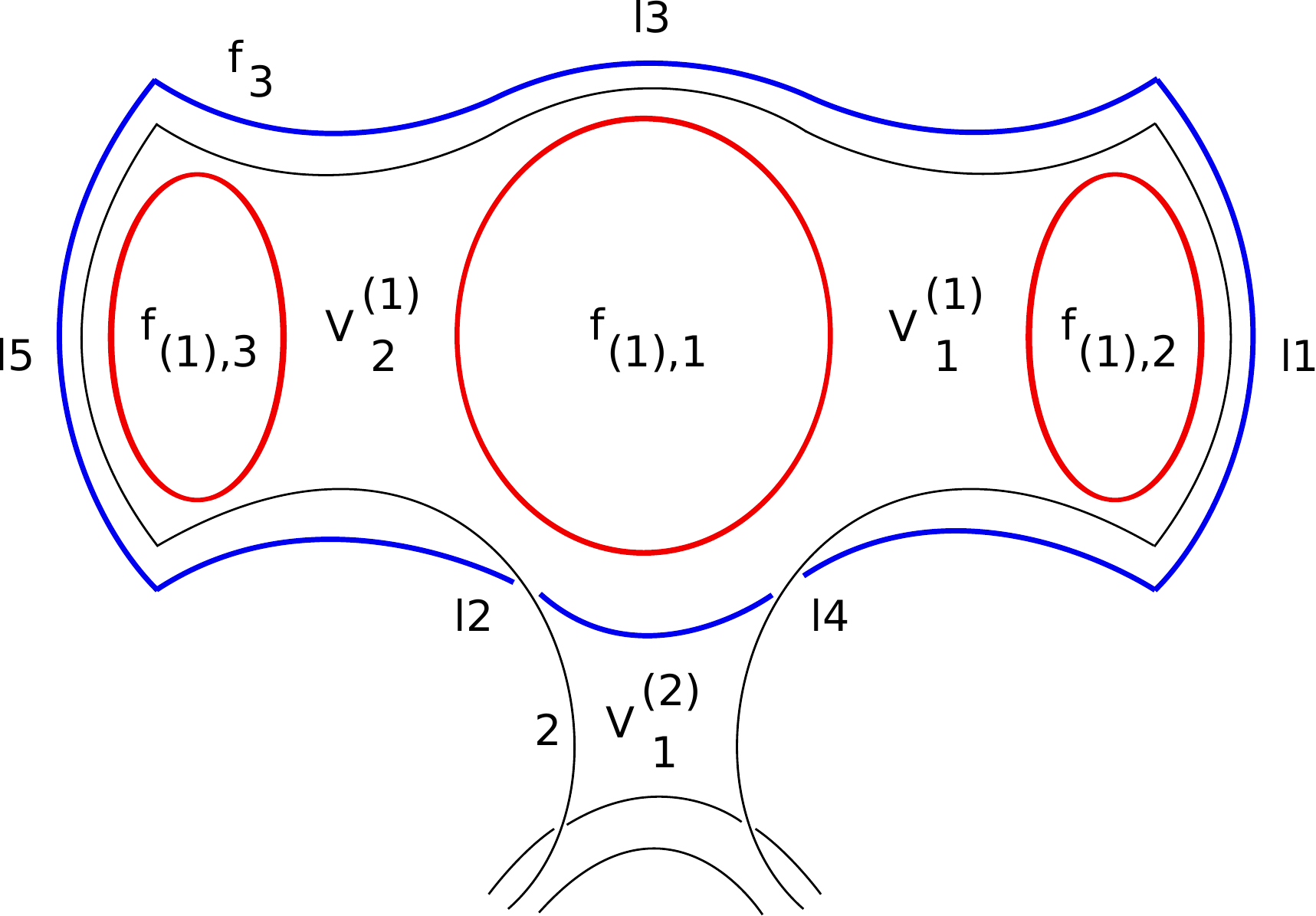} 
\vspace{0.8cm}
\caption{ {\small  A rank 3 graph with set  $\{V^{(1)}_1, V^{(1)}_2\}
\sqcup\{V^{(2)}_{1}\}$ of vertices, set  $\{l_1, \dots, l_5\}$ of  lines,
sets $\{f_{(1),1}, f_{(1),2}, f_{(1),3}\} \sqcup \{f_{3}\}$ of internal faces.  }} 
\label{fig:optim}
\end{minipage}
\end{figure}

Consider the graph $\cG$ of Figure \ref{fig:optim}. 
It is defined by a set $\cV$  of vertices decomposed in two disjoint subsets: 
$\{V^{(1)}_1, V^{(1)}_2\}$ which includes vertices with vertex 
kernel $V_{4;1}$ and $\{V^{(2)}_{1}\}$ identified by 
a kernel like $V_{4;2}$. 
The set $\cL$ of lines  includes $l_1,\dots,l_5$. 
We associate these lines with a scale index $i_{l_1}=15$, $i_{l_2}=12$,
$i_{l_3}=10$, $i_{l_4}=9$, $i_{l_5}=3$. 
The set of closed faces of $\cG$ decomposes into 
faces of color 1, namely $\{f_{(1),1}, f_{(1),2}, f_{(1),3}\}$ (in red),
and a face of color 3 denoted simply $\{f_3\}$ (in blue).

The optimization of the sums over internal momenta first requires
 to specify the index $i_f$ of each face such that
$i_{f}=\min_{l\in f} i_{l}$. This yields
\beq
i_{f_{(1),1}} = 9 \,,\qquad 
i_{f_{(1),2}} = 15\,,\qquad i_{f_{(1),3}} = 3\,,\qquad 
i_{f_{3}} =3 \,. 
\eeq
Each face $f$ has an index $i_f$ and momentum $p_{f}$ which 
is of  order $M^{i_{f}}$. According to our prescription, we can now arrange the $\hat\epsilon$ matrix as follow, from the highest $i_f$ to the lowest and by 
color blocks: 
\beq
\begin{array}{c|cc|c|}
 &V^{(1)}_1 & V^{(1)}_2 & V^{(2)}_1  \\ \hline
f_{(1),2} & 1 &  0  & 0 \\ 
f_{(1),1}  & 1 &1 & 0  \\ 
f_{(1),3}  &0  &1 &0\\ \hline 
f_{3} &0&0&0  
\end{array}
\label{tab}
\eeq
We start by $f_{(1),2}$ and count 
\bea
\varrho_{f_{(1),2}} = \sum_{V_k} \hat\epsilon_{ V_k, f_{(1),2} } = \hat\epsilon_{ V^{(1)}_1, f_{(1),2} } = 1 \,. 
\eea
then we erase the column $V^{(1)}_1$ and $f_{(1);2}$, and get the reduced matrix 
\beq
\begin{array}{c|c|c|}
  & V^{(1)}_2 & V^{(2)}_1  \\ \hline
f_{(1),1}  & 1 & 0  \\ 
f_{(1),3}  &1 &0\\ \hline 
f_{3} &0&0  
\end{array}
\label{tab}
\eeq
For $f_{(1),1}$, we count, using the above reduced matrix that we call 
again $\hat\epsilon$, 
\bea
\varrho_{f_{(1),1}} = \sum_{V_k} \hat\epsilon_{ V_k, f_{(1),1} } = \hat\epsilon_{ V^{(1)}_2, f_{(1),1} } = 1\,. 
\eea
Erasing the column $V^{(1)}_2$ leads to a trivial matrix and the procedure
stops. 
Here, forgetting any reference to scales, one concludes
that $\varrho(\cG)=1+1=2$. Note that in the above example 
$\varrho(\cG)$ is the rank of the $\hat\epsilon$ matrix but, more generally,
it might be not the case. 
We get the following optimal amplitude bound, for a constant $\kappa$,
\bea
A_{\cG;\bmu} \leq \kappa \big[\prod_{l\in \cL} M^{-2i_l} \big]
M^{ 9(1+2a) + 15(1+2a) + 3 +3   } \,.
\eea
It can be checked that, since $9(1+2a) + 15(1+2a) + 3 +3  = 30 + 24\times 2a $,
\bea
 \sum_{(i,k)} F_{\inter}(G^i_k) = 30 \,, 
\qquad 
\sum_{(i,k)} \varrho(G^i_k) = 24\,, 
\eea
with $F_{\inter}(G^{i\in [10,15]}_{k})=1$, $F_{\inter}(G^{i \in [4,9]}_{k})=2$,
$F_{\inter}(G^{i \in [1,3]}_{k})=4$, $\varrho(G^{i \in [10,15]}_1)=1$, 
 $\varrho(G^{i \in [11,12]}_2)=0$, and $\varrho(G^{i \in [1,9]}_2)=2$.


\end{document}